\documentclass[3p,times,procedia]{elsarticle}
\usepackage{nupha_ecrc}

\volume{00}

\firstpage{1}

\journalname{Nuclear Physics A}

\runauth{}

\jid{nupha}

\jnltitlelogo{Nuclear Physics A}

\usepackage{amssymb}

\usepackage{lineno}

\usepackage[figuresright]{rotating}
\usepackage{wrapfig}

\begin{document}

\begin{frontmatter}



\dochead{XXVIIth International Conference on Ultrarelativistic Nucleus-Nucleus Collisions\\ (Quark Matter 2018)}

\title{Global and local polarization of $\Lambda$ hyperons in Au+Au collisions at 200 GeV from STAR}


\author{Takafumi Niida (for the STAR Collaboration\footnotemark[1])}

\address{Wayne State University, 666 W. Hancock, Detroit, MI 48201}

\begin{abstract}
New results on global polarization of $\Lambda$ hyperons in Au+Au collisions at $\sqrt{s_{_{NN}}}=200$ GeV reveal non-zero signal on the order of a few tenths of a percent. Compared with lower energy results, the current measurement shows that the polarization decreases at higher collision energy, the trend being well reproduced by theoretical calculations. The polarization is found to be larger in more peripheral collisions, as well as in the in-plane than in the out-of-plane direction. The signal seems to depend on the asymmetry between positive and negative charge in each event, which may indicate an influence of the axial current induced by the magnetic field. In addition, the first measurement of a local polarization along the beam direction was performed. The results show a quadrupole modulation relative to the second-order event plane, as expected from the elliptic flow.
\end{abstract}


\end{frontmatter}

\footnotetext[1]{A list of members of the STAR Collaboration and acknowledgements can be found at the end of this issue.}

\section{Introduction}\label{sec:intro}
In non-central heavy-ion collisions, the created medium is supposed to have a large angular momentum transferred by two colliding nuclei. As it is shown in Ref.~\cite{Liang:2004ph,Voloshin:2004ha}, such an initial angular momentum would be partially transferred to the spin of produced particles and then the resulting global polarization would be experimentally detectable via hyperons through their parity-violating weak decays.
The STAR Collaboration performed the first global polarization measurements using $\Lambda$ hyperons at $\sqrt{s_{_{NN}}}=62.4$ and 200~GeV, where the signal was consistent with zero~\cite{Abelev:2007zk}.  More recently STAR has observed non-zero positive signal of $\Lambda$ global polarization in Au+Au collisions at $\sqrt{s_{_{NN}}}=7.7-39$ GeV~\cite{STAR:2017ckg} with a possible difference between $\Lambda$ and $\bar{\Lambda}$ that may be the effect of the initial magnetic field. 
The vorticity of the system indicated by the global polarization is an important piece in the picture of heavy-ion collisions and is possibly related to other observables such as directed and elliptic flow, and the chiral phenomena where the initial magnetic field plays an important role.

In this report, we present new results on the polarization of $\Lambda$ and $\bar{\Lambda}$ hyperons in Au+Au collisions at $\sqrt{s_{_{NN}}}=200$ GeV based on a recent data set which is about 150 times larger than that used in the previous analysis~\cite{Abelev:2007zk}.

\section{Analysis}\label{sec:ana}
The minimum bias data of Au+Au collisions at $\sqrt{s_{_{NN}}}=200$ GeV taken in 2010, 2011, and 2014 were analyzed in this study. The initial angular momentum points to the direction perpendicular to the reaction plane defined by the impact parameter vector and the beam direction. The first-order event plane, $\Psi_1$, was reconstructed using spectator fragments as an experimental estimate of the reaction plane. In the $\Lambda$ hyperon decay, daughter protons tend to be emitted in the direction of $\Lambda$'s spin. Therefore one can measure the spin polarization projected onto the initial angular momentum direction as follows~\cite{Abelev:2007zk}:
\begin{equation}
P_H = \frac{8}{\pi\alpha_H}\frac{\langle\sin(\Psi_1-\phi_p^{\ast})\rangle}{\rm Res(\Psi_1)},
\end{equation}
where $\alpha_H$ is the decay parameter of $\Lambda$ and $\bar{\Lambda}$ ($\alpha_{\Lambda}=-\alpha_{\bar{\Lambda}}=0.642\pm0.013$~\cite{PDG}), and $\phi_p^{\ast}$ is the azimuthal angle of the daughter proton in $\Lambda$ rest frame. The Res($\Psi_1$) is the resolution of the first-order event plane, which reaches a maximum of $\sim$0.4 in mid-central collisions. More details can be found in Ref.~\cite{Adam:2018ivw}.
 
\section{Results}
\begin{wrapfigure}[20]{r}[1mm]{0.40\linewidth}
\begin{center}\vspace{-22mm}
\includegraphics[width=\linewidth,angle=0,trim=40 0 30 0]{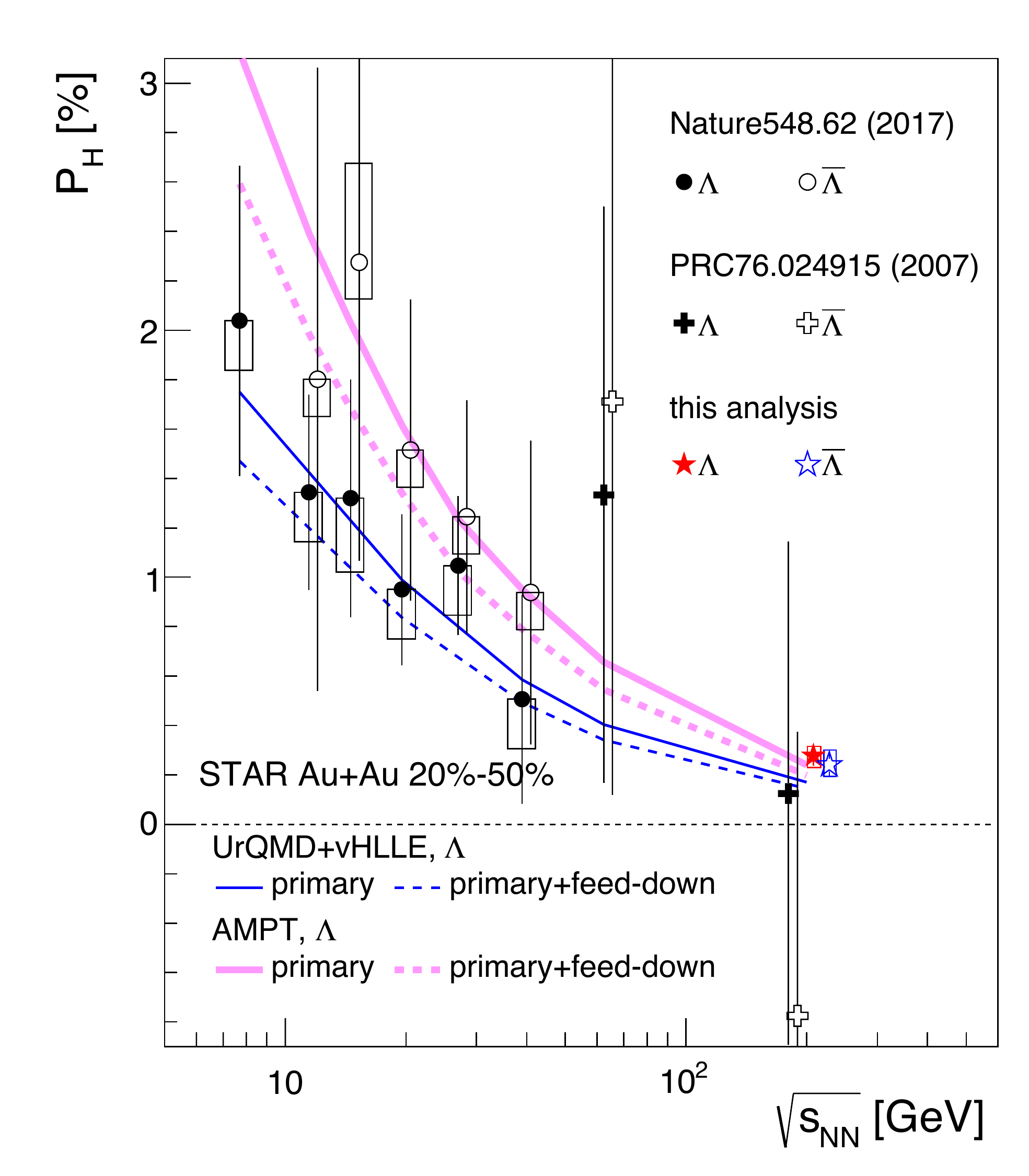}\vspace{-3mm}
\caption{Global polarization of $\Lambda$ and $\bar{\Lambda}$ as a function of collision energy for 20\%-50\% centrality in Au+Au collisions~\cite{Abelev:2007zk,STAR:2017ckg,Adam:2018ivw}. Calculations from hydrodynamic model~\cite{Karpenko:2016jyx} and AMPT model~\cite{Li:2017slc} with and without feed-down are compared.}\label{fig:BES}
\end{center}
\end{wrapfigure}
Figure~\ref{fig:BES} shows the collision energy dependence of $\Lambda$ global polarization for 20\%-50\% centrality in Au+Au collisions. New results for 200 GeV show non-zero positive signals for both $\Lambda$ and $\bar{\Lambda}$ and we confirm that the polarization decreases with increasing the collisions energy, by comparing with our results from lower energies. Hydrodynamic model with UrQMD initial condition~\cite{Karpenko:2016jyx} and a multi-phase transport (AMPT) model~\cite{Li:2017slc} reproduce the data well in the wide range of the collision energies $\sqrt{s_{_{NN}}}=$ 7.7--200~GeV.

\begin{figure}[hbt]\vspace{-5mm}
\begin{minipage}{0.5\hsize}
\begin{center}
\includegraphics[width=\textwidth,angle=0,trim=0 0 0 0]{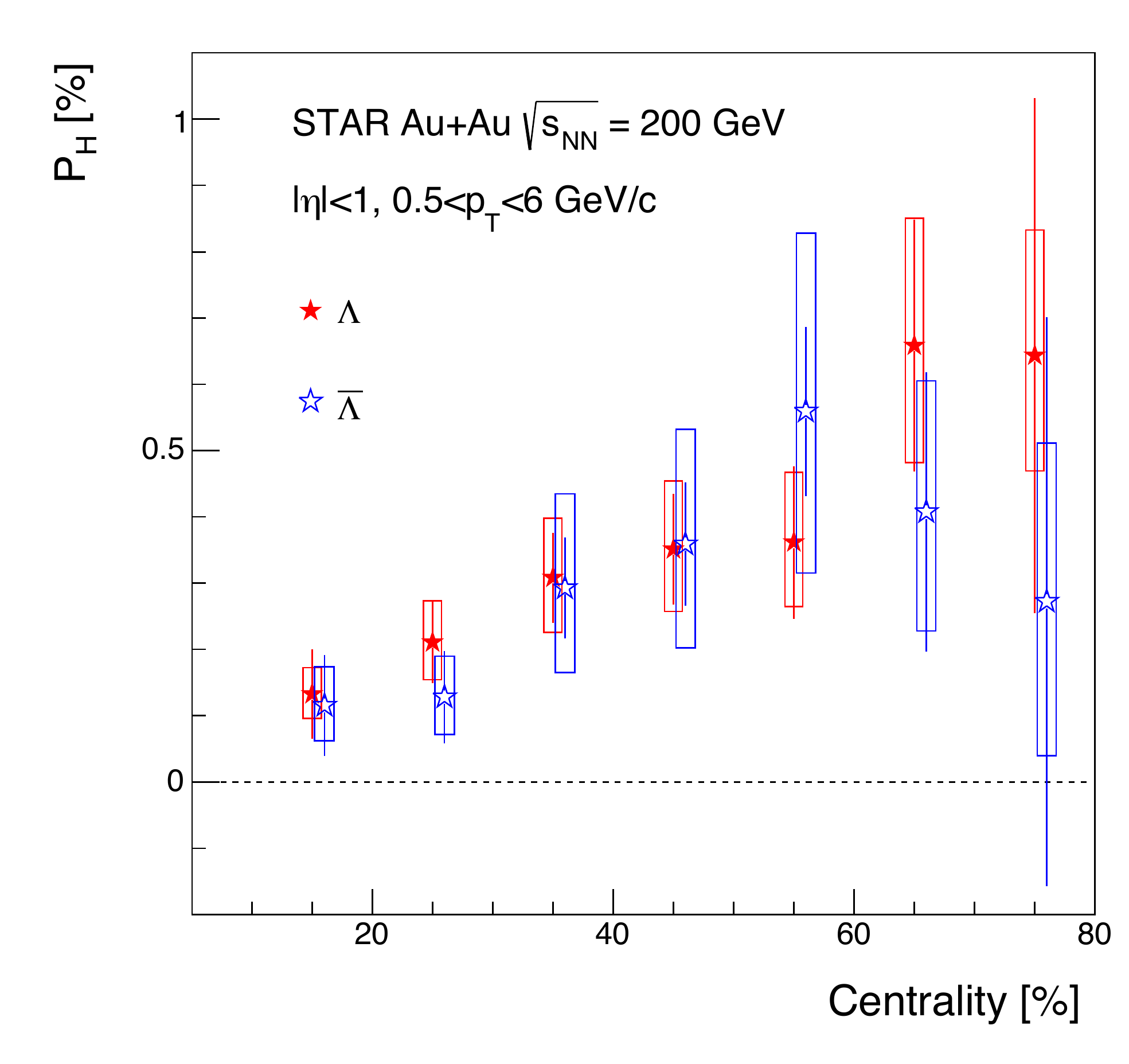}\vspace{-3mm}
\caption{Global polarization of $\Lambda$ and $\bar{\Lambda}$ as a function of centrality in Au+Au collisions at $\sqrt{s_{_{NN}}} = 200$ GeV. }\label{fig:cent}
\end{center}
\end{minipage}
\begin{minipage}{0.05\hsize}
\hspace{2mm}
\end{minipage}
\begin{minipage}{0.48\hsize}
\begin{center}\vspace{3mm}
\includegraphics[width=\textwidth,angle=0,trim=0 0 0 0]{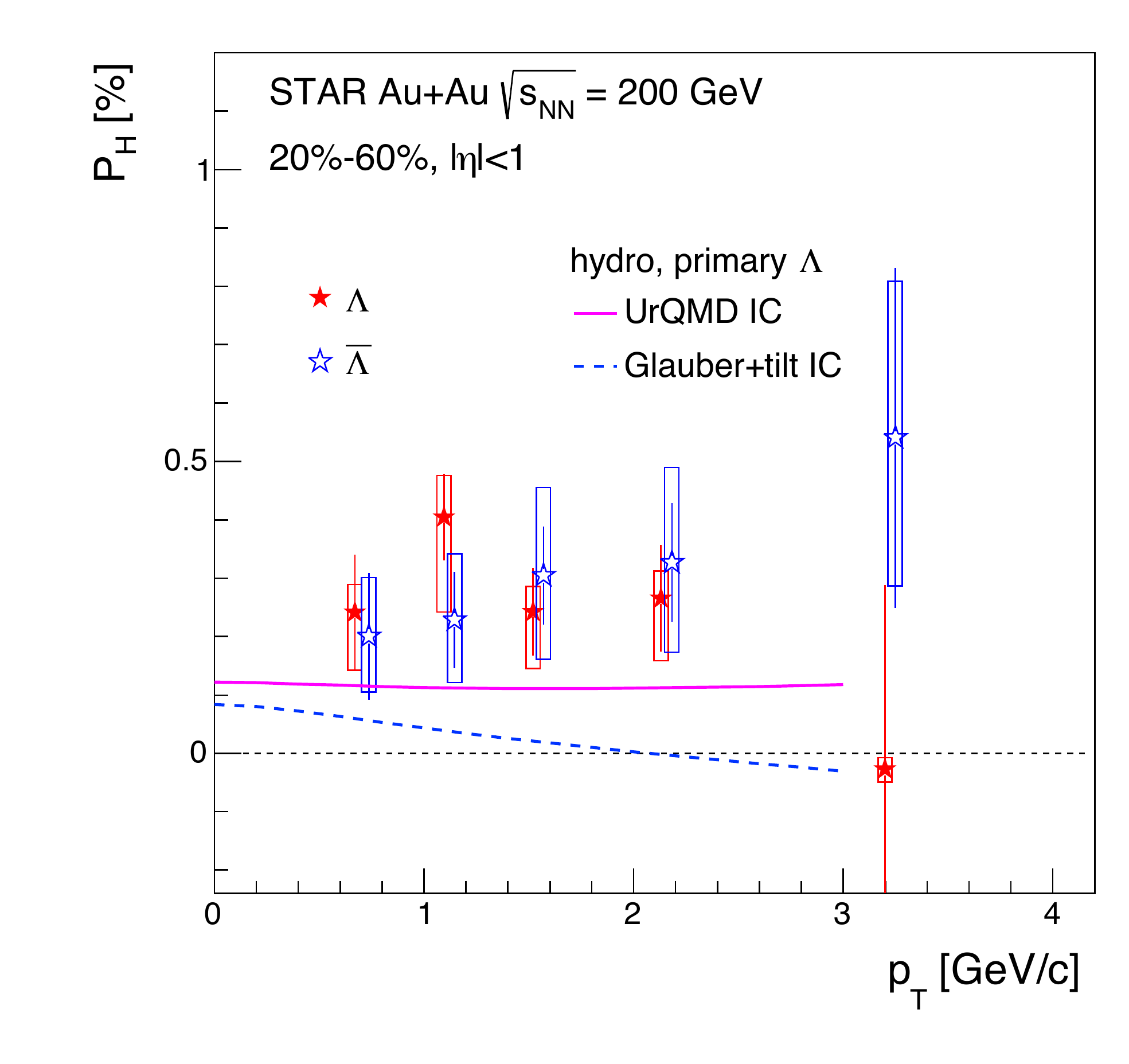}\vspace{-3mm}
\caption{Global polarization of $\Lambda$ and $\bar{\Lambda}$ as a function of $p_T$ for Au+Au 20\%-60\% central collisions~\cite{Adam:2018ivw}. Calculations from hydrodynamic model with two initial conditions: UrQMD and Glauber with tilted source~\cite{Karpenko:2016jyx} are compared.}\label{fig:pt}
\end{center}
\end{minipage}
\end{figure}

\begin{figure}[hbt]
\begin{minipage}{0.47\hsize}
\begin{center}
\includegraphics[width=\textwidth,angle=0,trim=170 100 140 100]{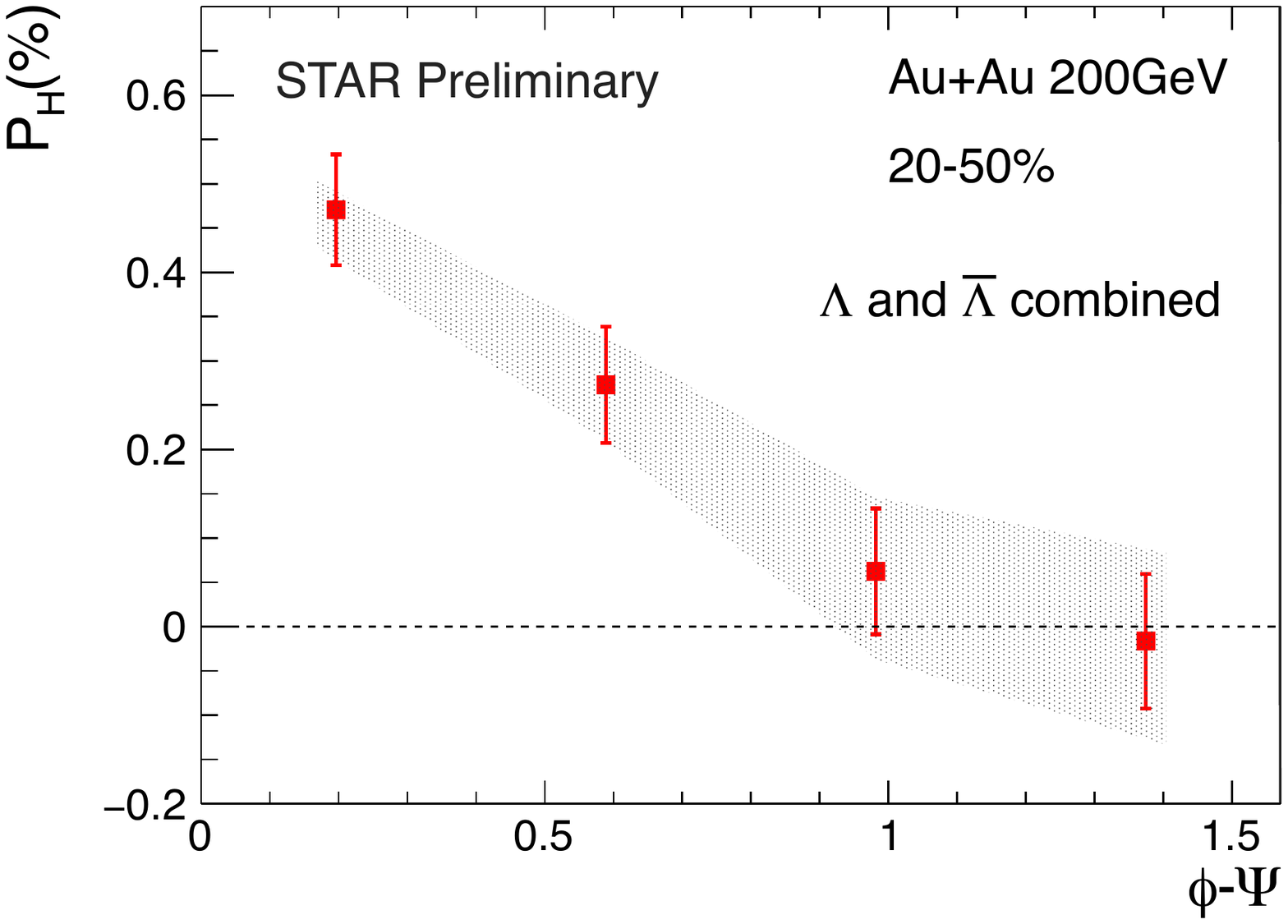}\vspace{-3mm}
\caption{Global polarization of $\Lambda$+$\bar{\Lambda}$ as a function of azimuthal angle $\phi$ relative to the first-order event plane for Au+Au 20\%-50\% central collisions. A shaded band shows systematic uncertainty.}\label{fig:RP}
\end{center}
\end{minipage}
\begin{minipage}{0.05\hsize}
\hspace{2mm}
\end{minipage}
\begin{minipage}{0.47\hsize}
\begin{center}\vspace{-2mm}
\includegraphics[width=0.98\textwidth,angle=0,trim=0 0 10 0]{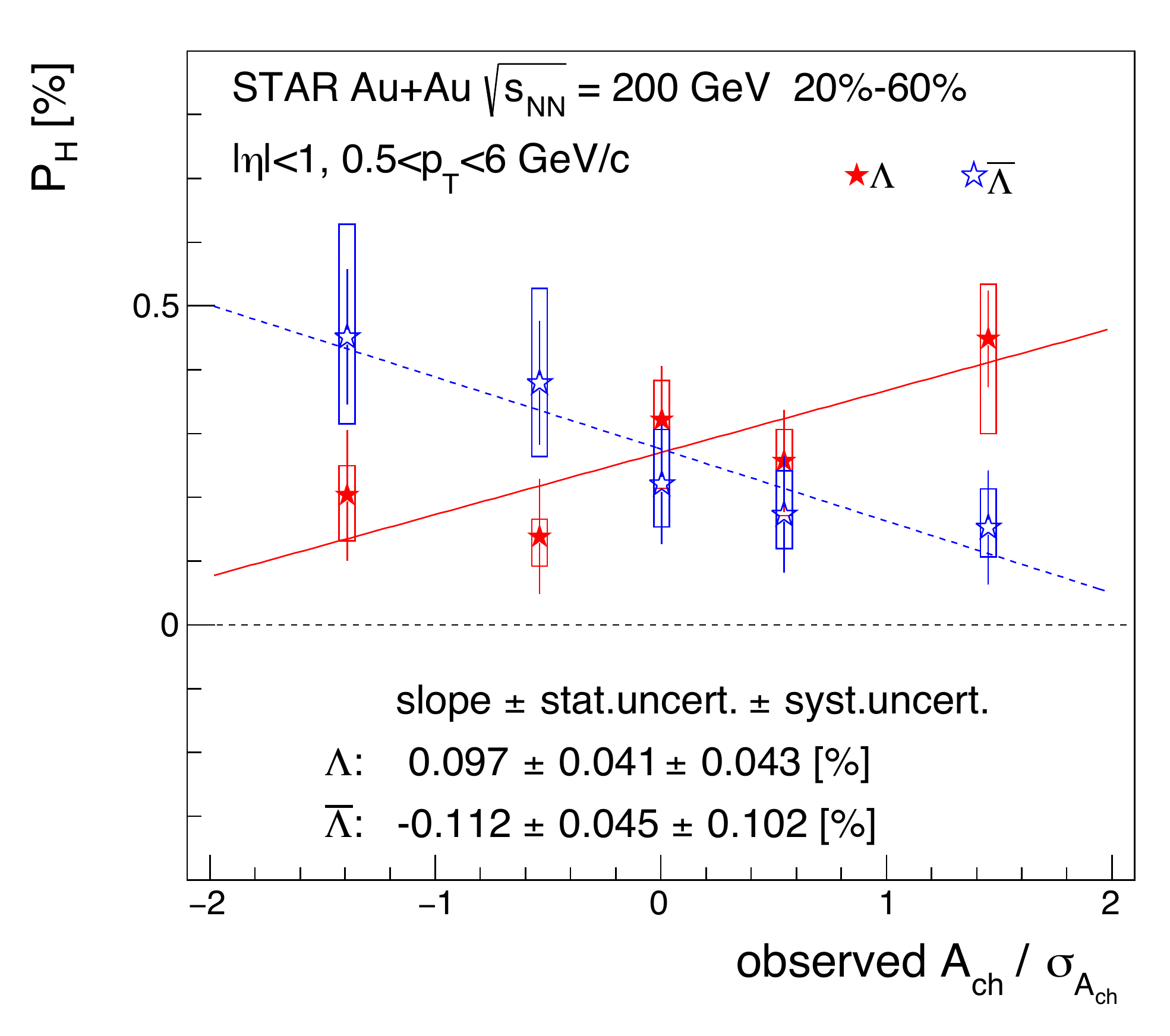}\vspace{-3mm}
\caption{Global polarization of $\Lambda$ and $\bar{\Lambda}$ as a function of observed charge asymmetry $A_{\rm ch}$ normalized by its RMS $\sigma_{A_{\rm ch}}$ for Au+Au 20\%-60\% central collisions at $\sqrt{s_{_{NN}}} = 200$ GeV. }\label{fig:ach}
\end{center}
\end{minipage}
\end{figure}

With the large dataset of 200 GeV, we further performed differential measurements such as the polarization versus centrality, hyperon's transverse momentum, and azimuthal angle.
Figure~\ref{fig:cent} shows centrality dependence of $\Lambda$ global polarization. The polarization increases in more peripheral collisions, as expected from an increase of thermal vorticity~\cite{Jiang:2016woz}. Figure~\ref{fig:pt} shows $p_T$ dependence of the polarization, comparing to hydrodynamic model with two different initial conditions. We found no significant dependence on $p_T$ within the uncertainties. The model calculations for primary $\Lambda$ underestimate the data, and the magnitudes and dependences slightly depend on the initial condition, but as in the data they also show very weak $p_T$ dependence. Figure~\ref{fig:RP} shows azimuthal angle dependence relative to the first-order event plane, and $\Lambda$ and $\bar{\Lambda}$ were combined to increase the statistical significance. The effect of the event plane resolution was corrected based on Ref.~\cite{Tang:2018qtu}. We found that the polarization is larger in the in-plane direction than in the out-of-plane, which is opposite to theoretical predictions, i.e. larger polarization in out-of-plane than in in-plane~\cite{Karpenko:2016jyx,Becattini:2013vja}.

The polarization might be explained by axial charge separation due to Chiral Vortical Effect~\cite{Baznat:2017jfj}. In addition to that, the axial current ${\bf J}_5$ can be created in the medium with non-zero vector chemical potential $\mu_v$ by the initial magnetic field ${\bf B}$ (${\bf J}_5\propto e\mu_v{\bf B}$, where $e$ is a particle electric charge), known as the Chiral Separation Effect~\cite{Kharzeev:2015znc}. Since the spins of particles in ${\bf J}_5$ are aligned to the direction of ${\bf B}$-field, there might be an additional contribution in the polarization from ${\bf J}_5$. In order to study this possible effect, we measured charge asymmetry
($A_{\rm ch}$) dependence of the polarization, assuming the relation $\mu_v / T \propto A_{\rm ch}$ where $A_{\rm ch}=\langle N_{+} - N_{-} \rangle / \langle N_{+} + N_{-} \rangle$ 
and $T$ is the temperature of the matter. Figure~\ref{fig:ach} presents $A_{\rm ch}$ dependence of the polarization.
The slopes extracted from a linear fit of this $A_{\rm ch}$ dependence seem to be different between $\Lambda$ and $\bar{\Lambda}$ ($\sim$2$\sigma$), which may indicate a possible contribution from the axial current induced by the magnetic field.

\begin{figure}[hbt]\vspace{-4mm}
\begin{minipage}{0.47\hsize}
\begin{center}
\includegraphics[width=0.9\textwidth,angle=0,trim=0 0 0 0]{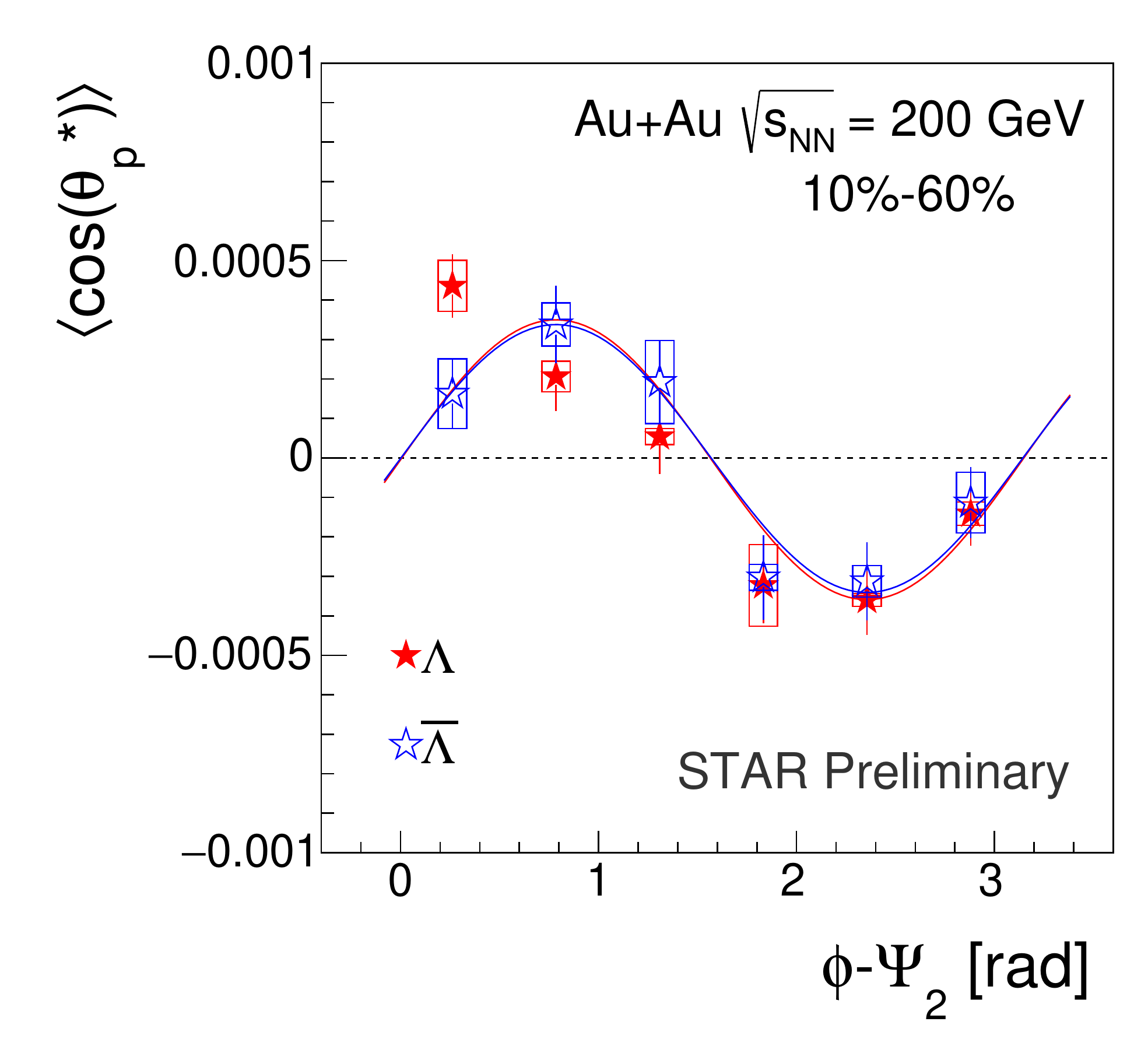}\vspace{-4mm}
\caption{ $\langle\cos\theta_p^{\ast}\rangle$ of $\Lambda$ and $\bar{\Lambda}$ as a function of hyperons' azimuthal angle $\phi$ relative to the second-order event plane $\Psi_2$ for 10\%-60\% centrality bin in Au+Au collisions at $\sqrt{s_{_{NN}}} = 200$ GeV. Resolution on $\Psi_2$ is not corrected here.}\label{fig:cos}
\end{center}
\end{minipage}
\begin{minipage}{0.05\hsize}
\hspace{2mm}
\end{minipage}
\begin{minipage}{0.47\hsize}
\begin{center}
\includegraphics[width=\textwidth,angle=0,trim=0 0 0 0]{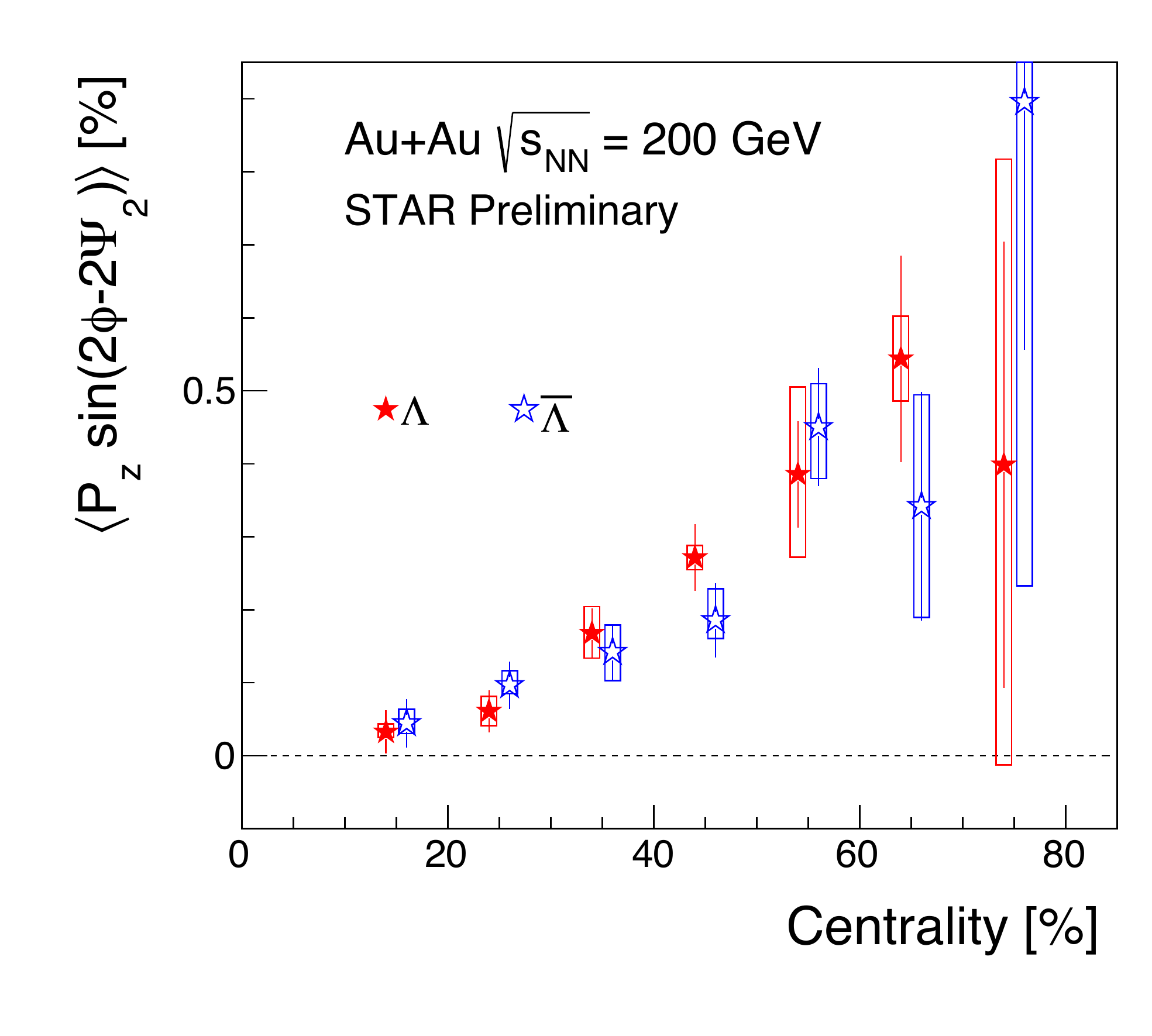}\vspace{-5mm}
\caption{Sine modulation of $\Lambda$ polarization along the beam direction relative to the second-order event plane as a function of centrality in Au+Au collisions at $\sqrt{s_{_{NN}}} = 200$ GeV.}\label{fig:pz2}
\end{center}
\end{minipage}
\end{figure}

The initial angular momentum of the medium provides an average direction of the vorticity and therefore the particle polarization, however the vorticity can be locally non-zero in general. 
As discussed in Refs.~\cite{Becattini:2017gcx,Voloshin:2017kqp}, a local vorticity along the beam direction may be created due to presence of elliptic flow. Similar to the global polarization, the polarization projected onto the beam direction, $P_z$, can be written as:
\begin{equation} 
P_z = \frac{\langle\cos\theta_p^{\ast}\rangle }{ \alpha_H\langle(\cos\theta_p^{\ast})^2\rangle },
\end{equation}
where $\theta_p^{\ast}$ is the polar angle of daughter proton in $\Lambda$ rest frame. The term $\langle(\cos\theta_p^{\ast})^2\rangle$ accounts for non-uniformity of detector in pseudorapidity and was calculated using the data (note that it becomes $1/3$ for a perfect detector). Figure~\ref{fig:cos} shows $\langle\cos\theta_p^{\ast}\rangle$ as a function of azimuthal angle relative to the second-order event plane. 
The data clearly shows a sine structure for both $\Lambda$ and $\bar{\Lambda}$, as expected from the elliptic flow. The second-order Fourier coefficient of the sine structure was measured as a function of centrality in Fig.~\ref{fig:pz2}. The coefficient increases in more peripheral collisions, similar to the centrality dependence of the elliptic flow. The sign of the coefficient is consistent with a naive expectation from the elliptic flow, but theoretical models predict the opposite sign~\cite{Becattini:2017gcx,Xia:2018tes}.

\section{Conclusions}
We present new results on $\Lambda$ and $\bar{\Lambda}$ global polarization in Au+Au collisions at $\sqrt{s_{_{NN}}}=200$ GeV. The significant improvement of the statistics using recent data allows to observe a non-zero positive signal at 200 GeV, which follows the global trend of the polarization versus collision energy. We found that the polarization becomes larger in more peripheral collisions, and larger in in-plane direction than in out-of-pane. The data also hints a weak dependence on charge asymmetry, which might be related to the axial current induced by the magnetic field. The local polarization along the beam direction was measured for the first time. The results shows a quadrupole modulation relative to the second-order event plane and this modulation becomes larger in more peripheral events, consistent with the expectation from the elliptic flow.

\section*{Acknowledgement}
This material is based upon work supported by the U.S. Department of Energy Office of Science, Office of Nuclear Physics under Award Number DE-FG02-92ER-40713.





\bibliographystyle{elsarticle-num}
\bibliography{ref_QM18proc}

\begin{thebibliography}{10}
\expandafter\ifx\csname url\endcsname\relax
  \def\url#1{\texttt{#1}}\fi
\expandafter\ifx\csname urlprefix\endcsname\relax\def\urlprefix{URL }\fi
\expandafter\ifx\csname href\endcsname\relax
  \def\href#1#2{#2} \def\path#1{#1}\fi

\bibitem{Liang:2004ph}
{Z.-T. Liang and X.-N. Wang}, Phys. Rev. Lett.  94 (2005) 102301, [Erratum:
  Phys. Rev. Lett.96,039901(2006)].
\newblock

\bibitem{Voloshin:2004ha}
S.~A. Voloshin, ~ \href {http://arxiv.org/abs/nucl-th/0410089}
  {\path{arXiv:nucl-th/0410089}}.

\bibitem{Abelev:2007zk}
{B. I. Abelev {\it et al.} (STAR Collaboration)}, Phys. Rev.  C76 (2007)
  024915, [Erratum: Phys. Rev.C95,no.3,039906(2017)].
\newblock

\bibitem{STAR:2017ckg}
{L. Adamczyk {\it et al.} (STAR Collaboration)}, Nature  548 (2017) 62--65.
\newblock

\bibitem{PDG}
{C. Patrignani {\it et al.} (Particle Data Group)}, Chin. Phys. C  40 (2016)
  100001.
\newblock

\bibitem{Adam:2018ivw}
{J. Adam {\it et al.} (STAR Collaboration)}, ~ \href
  {http://arxiv.org/abs/1805.04400} {\path{arXiv:1805.04400}}.

\bibitem{Karpenko:2016jyx}
{I. Karpenko and F. Becattini}, Eur. Phys. J.  C77 (2017) 213.
\newblock

\bibitem{Li:2017slc}
{H. Li, L.-G. Pang, Q. Wang and X.-L. Xia}, Phys. Rev.  C96 (2017) 054908.
\newblock

\bibitem{Jiang:2016woz}
{Y. Jiang, Z.-W. Lin, and J. Liao}, Phys. Rev.  C94 (2016) 044910, [Erratum:
  Phys. Rev.C95,no.4,049904(2017)].
\newblock

\bibitem{Tang:2018qtu}
{A. H. Tang, B. Tu, and C. S. Zhou} \href {http://arxiv.org/abs/1803.05777}
  {\path{arXiv:1803.05777}}.

\bibitem{Becattini:2013vja}
{F. Becattini, L. Csernai, and D. J. Wang}, Phys. Rev.  C88 (2013) 034905,
  [Erratum: Phys. Rev.C93,no.6,069901(2016)].
\newblock

\bibitem{Baznat:2017jfj}
{M. Baznat, K. Gudima, A. Sorin, and O. Teryaev}, Phys. Rev.  C97 (2018)
  041902.
\newblock

\bibitem{Kharzeev:2015znc}
{D. W. Kharzeev, J. Liao, S. A. Voloshin, and G. Wang}, Prog. Part. Nucl. Phys.
   88 (2016) 1.
\newblock

\bibitem{Becattini:2017gcx}
{F. Becattini, and Iu. Karpenko}, Phys. Rev. Lett.  120 (2018) 012302.
\newblock

\bibitem{Voloshin:2017kqp}
S.~A. Voloshin, [EPJ Web Conf.17,10700(2018)], \href
  {http://arxiv.org/abs/1710.08934} {\path{arXiv:1710.08934}}.

\bibitem{Xia:2018tes}
X.-L. Xia, H.~Li, Z.-B. Tang, Q.~Wang, ~ \href
  {http://arxiv.org/abs/1803.00867} {\path{arXiv:1803.00867}}.

\end{thebibliography}







\end{document}